\documentclass[fleqn,usenatbib]{mnras}

\usepackage{newtxtext,newtxmath}

\usepackage[T1]{fontenc}

\DeclareRobustCommand{\VAN}[3]{#2}
\let\VANthebibliography\thebibliography
\def\thebibliography{\DeclareRobustCommand{\VAN}[3]{##3}\VANthebibliography}

\usepackage{graphicx}	% Including figure files
\usepackage{amsmath}	% Advanced maths commands
\usepackage{amssymb}	% Extra maths symbols
\usepackage{multicol}
\usepackage{subcaption}

\newcommand{\intd}{{\rm d}}

\newcommand{\cm}{\,{\rm cm}}
\newcommand{\psec}{\,{\rm s}^{-1}}
\newcommand{\gram}{\,{\rm g}}

\newcommand{\rs}{\,r_{\rm s}}

\title[MHD disc winds]{Magnetically driven accretion disc winds: the role of gas thermodynamics and comparison to ultra-fast outflows}

\author[Wang, Bu \& Yuan]{Weixiao Wang,$^{1,2}$\thanks{E-mail:wxwang@shao.ac.cn (WW); dfbu@shao.ac.cn (DB); fyuan@shao.ac.cn (FY)} De-Fu Bu,$^{1}$ Feng Yuan$^{1}$\\
$^{1}$Shanghai Astronomical Observatory, Chinese Academy of Sciences, 80 Nandan Road, Shanghai 200030, People's Republic of China\\
$^{2}$University of Chinese Academy of Sciences, 19A Yuquan Road, Beijing 100049, People's Republic of China}

\pubyear{2022}

\begin{document}
\label{firstpage}
\pagerange{\pageref{firstpage}--\pageref{lastpage}}
\maketitle

\begin{abstract}
    Winds are commonly observed in luminous active galactic nuclei (AGNs). A plausible model of those winds is magnetohydrodynamic (MHD) disc winds. In the case of disc winds from a thin accretion disc, isothermal or adiabatic assumption is usually adopted in such MHD models. In this work we perform two-dimensional MHD simulations implementing different thermal treatments (isothermal, adiabatic and radiative) to study their effects on winds from a thin accretion disc. We find that both the isothermal model and the adiabatic model overestimate the temperature, underestimate the power of disc winds, and cannot predict the local structure of the winds, compared to the results obtained by solving the energy equation with radiative cooling and heating. Based on the model with radiative cooling and heating, the ionization parameter, the column density and the velocity of the disc winds have been compared to the observed ultrafast outflows (UFOs). We find that in our simulations the UFOs can only be produced inside hundreds of Schwarzschild radius. At much larger radii, no UFOs are found. Thus, the pure MHD winds cannot interpret all the observed UFOs.
\end{abstract}


%%%%%%%%%%%%%%%%% BODY OF PAPER %%%%%%%%%%%%%%%%%%
\section{Introduction}

Winds from active galactic nuclei (hereafter AGNs) are ubiquitous. Recently observed absorbers which are called ultra-fast outflows (UFOs) are highly ionized, with the ionization parameter $\xi$ in the range of $\rm{log}(\xi/erg\,s^{-1}\,cm)\sim 3-6$, and mildly relativistic, with velocity $v>0.1\,c$ ($c$ is the speed of light) \citep[e.g.][]{Pounds2003,Reeves2003,Tombesi2010,Gofford2013}. The UFOs have a hydrogen-equivalent column density in the range of ${\rm log}(N_{\rm H}/cm^{-2})\sim 22-24$. The estimated location of UFOs is about hundreds to thousands of Schwarzschild radius ($\rs$) from the central black hole \citep[e.g.][]{Reeves2010,Tombesi2010,Tombesi2012,Tombesi2013,Gofford2015}. It is believed that UFOs are winds launched from a standard thin disc \citep{ShakuraSunyaev1973} around the central black hole \citep[e.g.][]{KingPounds2003,ProgaKallman2004,Ohsuga2009,Fukumura2015,Nomura2016,NomuraOhsuga2017}.

In the aspect of theories, the presence of wind can affect the dynamics and spectrum of the accretion flow. For example, the angular momentum of an accretion disc can be removed by winds, which drives gas fall onto the central object \citep[e.g.][]{Konigl1989,StoneNorman1994,BaiStone2013,LiBegelman2014,ZhuStone2018,LiCao2019}. Winds can also influence the density and temperature profiles of an accretion disc, and subsequently change the emitted spectrum \citep[e.g.][]{Li2019,Yuan2003}. Moreover, it is generally believed that winds from AGNs can affect the evolution of its host galaxy via the ``AGN feedback'' process \citep[e.g.][]{Fabian2012,KingPounds2015, Chen2022, He2022}. Thus, the wind production and propagation in and from AGNs are of great interest.

It is well known for a long time that magnetic field plays a prominent role in driving winds. When the poloidal magnetic field at the disc surface is inclined at an adequately large angle from the rotation axis, winds can be driven by magneto-centrifugal force \citep[e.g.][]{BlandfordPayne1982,CaoSpruit1994,ContopoulosLovelace1994}. If its toroidal component is dominant, winds can be driven by the magnetic pressure gradient force \citep[e.g.][]{LyndenBell1996,Kato2004,Yuan2015}. Alternatively, radiatively driven mechanism \citep[e.g.][]{Proga2000,ProgaKallman2004} and thermally driven mechanism \citep[e.g.][]{Begelman1983,Dyda2017,WatersProga2018} are plausible for launching disc winds. In this paper, we only focus on the magnetically driven mechanism.

From the view of thermodynamics of disc winds, early analytical literature of MHD disc winds had assumed an isothermal/adiabatic process or a polytropic process with a parametrized index \citep[e.g.][]{ContopoulosLovelace1994,CaoSpruit1994,KudohShibata1997}. It results in a quite simple description of the thermal properties. They also assumed self-similar solutions to make it feasible to analytically solve the set of MHD equations. Even in semi-analytical work the treatments for the thermal properties are limited \citep[e.g.][]{LiCao2021}. When an energy equation including radiative cooling and heating is considered, the MHD equations cannot be normalized to a dimensionless form; self-similar solutions are no longer valid. Therefore, numerical studies are invoked to self-consistently obtain the wind solutions when considering an appropriate energy equation.

In the case of winds launched from a cold thin disc, technical difficulties in numerically simulating a geometrically thin accretion flow have limited studies and understandings of the disc winds. Lots of efforts have been made to self-consistently solve the accretion process by global simulations \citep[e.g.][]{ZhuStone2018,Mishra2020}. However, the computational domain in these simulations is too narrow to obtain complete wind solutions. In addition, the scale height $H/r$ of a typical standard thin disc with characteristic parameters\footnote{Black hole mass $M_{\rm BH}=10^8M_{\sun}$, the prescription of viscosity $\alpha=0.1$ and mass accretion rate $\dot{M}_{\rm BH}=0.1\dot{M}_{\rm Edd}$.} in a luminous AGN is in the order of $10^{-3}$, which has hitherto never been approached in global simulations. In the case of minimal scale height, the disc can be regarded as a boundary without being resolved if it is assumed that the presence of winds do not influence the disc structure.

In this work, we will show the difference of wind properties between the isothermal model, the adiabatic model and the radiative model (as our fiducial model). The purpose is to investigate whether the isothermal or adiabatic assumption on the disc winds are valid and adequate. 

Apart from the thermodynamics of MHD disc winds, their observable properties are of great interest. The MHD wind model has been employed to explain the observed UFOs. \citet{Fukumura2015,Fukumura2018} utilize the self-similar MHD wind solution to interpret the UFOs in nearby quasars PG 1211+143 and PDS 456. The authors adjust wind model parameters to fit the observed spectra. They find that for proper parameters, the self-similar MHD winds can well produce the observed UFOs. Therefore, they believe that the UFOs are MHD driven accretion disc winds. Nevertheless, the self-similar solutions oversimplify the wind model. Whether the MHD winds are radially self-similar is unclear. In this case, numerical studies of the disc winds considering radiations are still necessary for the realistic case especially where self-similarity is violated.

According to these considerations, the first aim of the present paper is to study the effects of different thermodynamics treatment on the properties of winds from a thin disc. The second aim is to study whether UFOs can be launched by magnetic field. The paper is organized as follows. Section \ref{sec:methods} provides basic equations and assumptions of the model.  Section \ref{sec:simulation} describes numerical setup. The results are presented in Section \ref{sec:results}. We summarize our work in Section \ref{sec:sum}.

\section{Methods}\label{sec:methods}

\subsection{Governing Equations}
\label{sec:eqns}

Cold accretion disc in luminous AGNs is geometrically thin. Technically, it is arduous to resolve the thin accretion disc in global simulations.  In this paper, we simulate winds launching from the surface of a thin disc. Same as the way that \citet{Ustyugova1999}, \citet{Proga2000}, \citet{Nomura2016} and \citet{NomuraOhsuga2017} implement the disc, the main body of the thin disc is not resolved and not included in the computational domain; rather, the surface of the accretion disc is treated as the boundary of our simulation. Spherical coordinates $(r, \theta, \phi)$ are employed and we only simulate the region above the midplane. The plane of $\theta=90^\circ$ corresponds to the disc surface. Following \citet{Nomura2016} and \citet{NomuraOhsuga2017}, we assume that the black hole is located at the location $\theta=180^\circ, r\cos \theta = -3.1\varepsilon \rs $, where $\varepsilon$ is the Eddington ratio of the AGN. Observations imply that a compact hot corona should exist sandwiching the disc in the innermost region. We assume that the corona is a point source and is located at the origin ($r=0$).

The dynamics of the winds are governed by the following equations in general,
\begin{equation}\label{eqn:rho}
 \frac{\partial\rho}{\partial t} + \nabla\cdot(\rho{\bf v}) = 0,
\end{equation}
\begin{equation}\label{eqn:v1}
 \rho \frac{\partial v_r}{\partial t} + \rho ({\bf v} \cdot \nabla) v_r = -\frac{\partial P}{\partial r}  +
 \rho g_r + f_{\rm {B,r}} + \rho (\frac{v_\theta^2}{r}+\frac{v_\phi^2}{r})
\end{equation}
\begin{equation}\label{eqn:v2}
 \rho \frac{\partial v_\theta}{\partial t} + \rho ({\bf v} \cdot \nabla) v_\theta = -\frac{\partial P}{\partial \theta}  +
 \rho g_\theta + f_{\rm {B,\theta}} - \rho (\frac{v_\theta v_r}{r}-\frac{v_\phi^2}{r\tan \theta})
\end{equation}
\begin{equation}\label{eqn:v3}
 \rho \frac{\partial v_\phi}{\partial t} + \rho ({\bf v} \cdot \nabla) v_\phi = f_{\rm {B,\phi}} - \rho (\frac{v_\theta v_r}{r}-\frac{v_\phi v_\theta}{r\tan \theta})
\end{equation}
\begin{equation}\label{eqn:energy}
 \frac{\partial (\rho e)}{\partial t} + \nabla\cdot(\rho e{\bf v}) = -P\nabla \cdot {\bf v} + \rho \bf {\mathcal {L}}
\end{equation}
\begin{equation}\label{eqn:mag}
	\frac{\partial \bf B}{\partial t} = \nabla\times(\bf{v}\times\bf{B})
\end{equation}
where $\rho$ is mass density, $\bf v$ is the velocity and $e$ is the gas specific internal energy, $P$ is the gas pressure, ${\bf f}_{\rm B}= (f_{\rm {B,r}},f_{\rm {B,\theta}},f_{\rm {B,\phi}})$ is the Lorentz force per unit volume, ${\bf g}=(g_r, g_\theta)$ is the gravity whose $\theta$-component arises because the black hole is not located at the origin and $\mathcal{L}$ is the net heating/cooling function. A Newtonian potential is utilized. The equation of state for perfect gas is adopted: $P=(\gamma-1)\rho e$ with $\gamma=5/3$.

Without loss of generality, the energy equation is represented in the form of Equation \ref{eqn:energy}. When $\mathcal{L}=0$, it describes adiabatic flow; if Equation \ref{eqn:energy} is not solved but replaced by $e\propto\rho$, it displays the case of isothermal process.

\subsection{Radiative heating and cooling}

The X-ray radiation from the corona heats, cools and photoionizes gas in winds through radiative processes of Compton scattering and photoionization. Additionally, winds lose energy via bremsstrahlung, line cooling and recombination cooling. Here we use exactly the same cooling function as in \citet[]{Blondin1994}. In Equation (\ref{eqn:energy}) the net cooling rate is written as
\begin{equation}
\rho\mathcal{L}=n^2(\Gamma_{\rm Compton}+\Gamma_{\rm X}-\Lambda),
\end{equation}
where $\Gamma_{\rm Compton}$ is the heating or cooling rate of Compton scattering
\begin{equation}
\Gamma_{\rm Compton} = 8.9 \times 10^{-36}\xi(T_{\rm X}-4T),
\end{equation}
$\Gamma_{\rm X}$ is the photoionization heating/recombination cooling,
\begin{equation}
\Gamma_{\rm X} = 1.5\times 10^{-21}\xi^{1/4}T^{-1/2}(1-T/T_{\rm X}),
\end{equation}
and $\Lambda$ is the cooling rate of bremsstrahlung and line cooling
\begin{equation}
\begin{split}
\Lambda = &3.3\times 10^{-27}T^{1/2}+[1.7\times 10^{-18}\\
&{\rm exp}(-1.3\times 10^{5}/T)\xi^{-1}T^{-1/2}+10^{-24}]\delta.
\end{split}
\end{equation}\\
In the above equations, $T_{\rm X}$ is the characteristic temperature of the X-ray radiation and $\delta$ is a parameter to represent optically thin or thick atmosphere. $T_{\rm X}$ is set to $10^8\rm K$ corresponding to $\rm 10\,keV$ bremsstrahlung radiation from the central corona. Assuming the wind is optically thin to its own  radiation, we have $\delta=1$.

\section{Simulations}\label{sec:simulation}
We solve the Equations (\ref{eqn:rho})-(\ref{eqn:mag}) using the ZEUS-3D code \citep{Clarke1996}. ZEUS-3D is paralleled by OpenMP and solves the fluid equations using the finite-difference method. We adopt the same computational domain and grid spacing as those in \citet{Nomura2016}: $30\rs \leq r \leq 1500\rs$ and $0 \leq \theta \leq \pi/2$. The resolution is $N_r\times N_\theta =100\times140$. In order to well resolve the inner radial region, we adopt logarithmic grid with $(\Delta r)_{i+1}/(\Delta r)_i=1.05$ in the radial direction. In the $\theta$ direction, the resolution increases towards the equatorial plane with $(\Delta \theta)_{j+1}/(\Delta \theta)_j=1.066$.

\subsection{Accretion disc treatment}\label{subsec:dsktreat}
In this paper, we simulate winds launching from the surface of a thin disc. The surface of the disc is implemented in the first layer of grids of the computational domain above the equatorial plane (with $z=0$). The black hole is located below the surface of accretion disc and following \citet{NomuraOhsuga2017} we put the black hole at the location ($r=z_0=-3.1\varepsilon \rs$, $\theta=\pi$). $z_0$ corresponds to the scale height of the accretion disc at the inner radial boundary, which is $30\rs$. According to the standard thin disc model \citep{ShakuraSunyaev1973}, the azimuthal velocity equals to the Keplerian velocity at every radius. We assume that the density of the surface of the disc does not vary with time. In this sense, the mass taken away by wind will be immediately replenished. This setting is similar as that in \citet[]{Nomura2016}. In that paper, the authors assume that the density of the surface of the disc at the midplane is constant with both time and radius. In our present paper, we assume that the density of the surface of the accretion disc varies with radius, but not with time. In order to set a reasonable non-evolving accretion disc surface density, we employ the accretion disc theory. We set it according to the standard thin accretion disc value given by \citep[e.g.][]{NomuraOhsuga2017,Kato2008book},
\begin{align}
&\rho_{\theta=\pi/2}\,(r) = \nonumber\\
&\begin{cases}
5.24*10^{-4}(M_{\rm BH}/M_\odot)^{-1}(\varepsilon/\eta)^{-2}(r/\rs)^{3/2}
&r \leq r_{\rm crit},\\
4.66(M_{\rm BH}/M_\odot)^{-7/10}(\varepsilon/\eta)^{2/5}(r/\rs)^{-33/20}
&r>r_{\rm crit},
\end{cases}
\end{align}\\
where $r_{\rm crit}=18(M_{\rm BH}/M_\odot)^{2/21}(\varepsilon/\eta)^{16/21}\rs$ is the critical radius. Inside $r_{\rm crit}$, radiation pressure dominates gas pressure; outside $r_{\rm crit}$, gas pressure dominates radiation pressure. According to the parameters adopted in the present paper, $r_{\rm crit} \sim 154\rs$. $\eta$ is the radiative efficiency of the accretion disc. We note that the radial velocity at the midplane is not calculated according to the thin disc theory. Initially, we set it to be 0. It can evolve with time. Therefore, the accretion rate of the midplane disc is not a constant with time. In our fiducial model, we set $\varepsilon=0.1$. This does not mean that the accretion disc has an accretion rate of 0.1 times Eddington rate. We just use Equation (11) to set a reasonable density of the surface of a thin disc. The temperature profile is,
\begin{equation}
T_{\theta=\pi/2}\,(r) = T_{\rm in}\bigg(\frac{r}{r_{\rm in}}\bigg)^{-3/4}
\end{equation}\\
where $r_{\rm in}$ is the inner radius of the disc and is set to be $3\rs$, $T_{\rm in}$ is the effective temperature at $r_{\rm in}$,  which is set to satisfy $L_{\rm D}=\int_{\rm in}^{\rm out}2\pi r\sigma T^4 {\rm d} r$, where $\sigma$ is the Stefan-Boltzmann coefficient. Under the assumption made in Section 2.3, the outer radius $r_{\rm out}$ can be derived from $T_{\rm in}(r_{\rm out}/r_{\rm in})^{-3/4}=3\times10^3{\rm K}$. The evolution of disc is neglected. The density and the temperature of the disc are fixed, but we allow the velocities ($v_r, v_{\theta}, v_{\phi}$) and fields ($b_r, b_{\theta}, b_{\phi}$) to float.

\subsection{Initial and boundary conditions}\label{subsec:bc}
The gas in the computational domain is initially set to be vertically in hydro-static equilibrium without radiative force. Then the initial density distribution above the surface of the thin disc reads,
\begin{equation}
\rho_0\,(r,\theta)= \rho_{\theta=\pi/2}\,(r)\,{\rm{exp}}\left(-\frac{GM_{\rm BH}}{2c_s^2r\tan^2\theta}\right).
\end{equation}\\
Here $G$ is the gravitational constant and $c_s$ is the sound speed. Since the density in the vertical direction is exponentially decreasing, the ratio of density at equatorial plane to density at high latitude can increase with height and approaches infinity. In numerical simulations if the density discrepancy in the computational domain is too large, the simulation is usually quite hard to run. In order to overcome this problem, we set a density floor to be $\rho_{\rm floor}=1.49\times10^{-20}\gram\cm^{-3}$. Note that the value of density floor is 11 orders of magnitude smaller than the density at the equatorial plane,  and therefore, it has negligibly small effects on our results. Initially, the gas has Keplerian rotational velocity. The initial radial and $\theta$ directions velocities are set to be zero. So the initial velocity is $v_0=(0,0,\frac{GMr\sin\theta}{\sqrt{z_0^2+r^2+2rz_0\cos\theta}})$. The initial temperature is vertically isothermal so that $T_0(r,\theta)=T_{\theta=\pi/2}(r\sin\theta)$. The initial internal energy density can be derived by using $\rho_0(r,\theta)$ and $T_0(r,\theta)$. Additionally, initial ionization parameter is needed to determine the opacity at the first time step of simulations, and the initial opacity is needed to determine the ionization X-ray flux at the fist time step. Due to the low density of background, the gas is assumed to be initially photoionized by X-ray without attenuation.

\begin{figure}
	\includegraphics[width=\columnwidth]{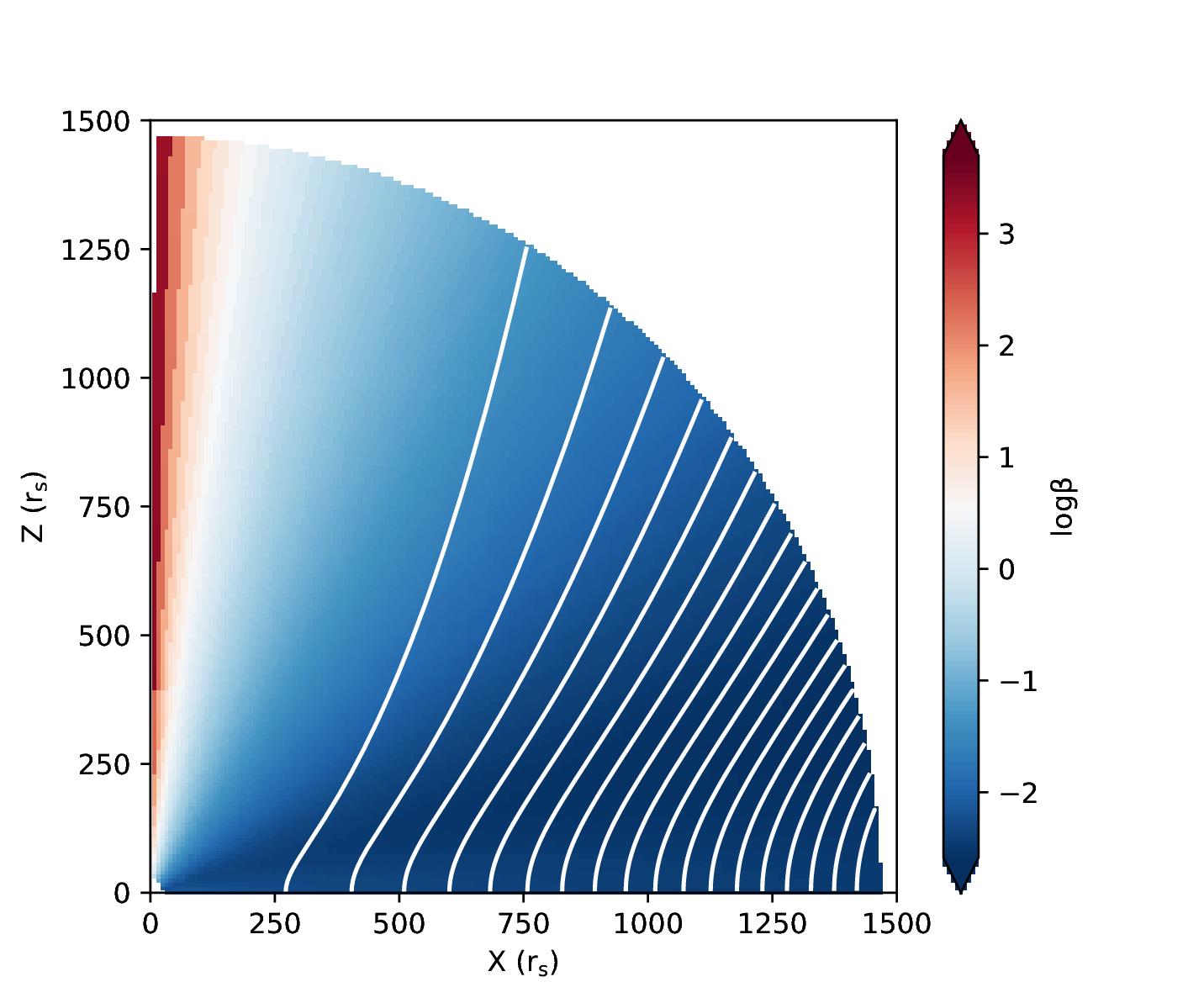}
    \caption{Colormap shows the logarithmic plasma beta and white curves represent the magnetic field lines. The initial $\beta$ is about $10^8$ on the disc surface, i.e. the first layer of grid above the equator.}
    \label{fig:initialBfield}
\end{figure}

The initial magnetic field is defined by using the vector potential below \citep[][]{Zanni2007},
\begin{equation}
A_\phi(r,\theta)=B_0(r \sin\theta)^{3/4}\frac{m^{5/4}}{(m^2+\tan^{-2}\theta)^{5/8}}.
\end{equation}
The parameter $B_0$ determines the strength of the initial magnetic field. The inclination angle of the initial magnetic field lines with respect to the equatorial plane is determined by the parameter $m$. In this paper, we set $m=0.4$. The initial plasma $\beta$, defined as the ratio of gas pressure to magnetic pressure, is shown in Fig.~\ref{fig:initialBfield}.

We impose outflow boundary conditions at the inner and outer radial boundaries. Axi-symmetry boundary condition are set at $\theta=0$. At $\theta=\pi/2$, we use reflecting boundary conditions.

\subsection{Models and parameters}\label{subsec:models}
We assume the mass of a non-rotating black hole $M_{\rm BH} = 10^8 M_\odot$ and the radiative efficiency of the accretion disc is $\eta=0.06$. The disc luminosity is set to be 0.1 times the Eddington luminosity.  In this paper, we do not simulate the accretion process. The effects of winds on the accretion disc are neglected. In reality, in an accretion system, the presence of winds can affect the accretion rate of the disc. In the hydrodynamic simulations, \citet[]{Nomura2020} have examined the effects of mass loss on the disc. The authors utilized iterative method to adjust the accretion rate. The iteration which can reach convergence is a plausible method. However, in the MHD case, the situation becomes complicated, since the magnetic field evolves with the fluid. It is difficult to determine a self-consistent magnetic field when one does iteration.In the present MHD simulations, we do not use the iterative method to adjust the accretion rate.

In order to compare the different thermodynamics influencing the disc winds, we consider adiabatic and isothermal processes as well. For the isothermal model, the internal energy does not evolve so that the Equation \ref{eqn:energy} is eliminated. For the adiabatic model, in Equation \ref{eqn:energy} the cooling function $\mathcal{L}$ is set to zero. Without radiation, the ionization parameter can not be defined. The set of equations remain closed. Models with referred parameters are listed in Table~\ref{tab:models}.

\begin{table}
	\centering
	\caption{All models have initial plasma beta $\beta_0=10$ on the first cell of grids.\\
	$^*$ denotes the fiducial model.}
	\label{tab:models}
	\begin{tabular}{llc}
	\hline
	Model Name & Thermal process & temperature \\
	\hline
    FID$^*$ & Blondin's $\mathcal{L}$ & self-consistently derived \\
    ADB & Adiabatic & self-consistently derived \\
    ISO6 & Globally isothermal & $10^6$K ($c_s\approx 111 {\rm km}\psec$) \\
    ISO7 & Globally isothermal & $10^7$K ($c_s\approx 351 {\rm km}\psec$) \\
    \hline
	\end{tabular}
\end{table}

%%%%%%%%%%%%%%%%%%%%% RESULTS %%%%%%%%%%%%%%%%
\section{results}\label{sec:results}
In this section, we present simulation results and comparison with observations. It is organized as follows. Section \ref{subsec:tave} describes the time-averaged quantities; real winds are identified in Section \ref{subsec:realwind}; in Section \ref{subsec:windprop} we show the detailed properties of real winds and demonstrate the dependence on different thermodynamics; Section \ref{subsec:fid_mech} describes the mechanisms of driving disc winds for the fiducial model FID; Section \ref{subsec:fid_obs} compares the observable features of disc winds with the UFOs.

\subsection{Time-averaged quantities}\label{subsec:tave}

\begin{figure}
	\includegraphics[width=\columnwidth]{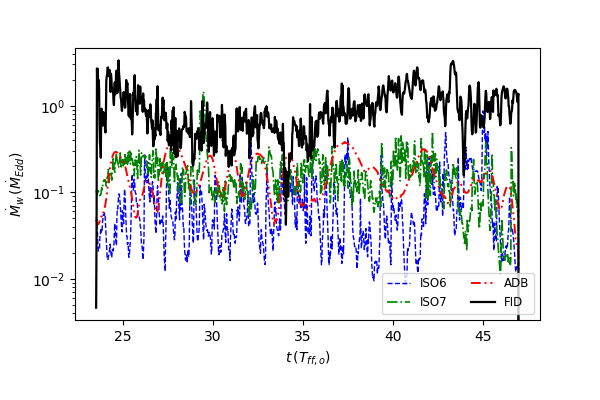}
    \caption{Evolution of mass loss rate at the outer radial boundary in unit of Eddington limit $\dot{M}_{\rm Edd} = \dot{L}_{\rm Edd}/\eta/c^2$, where $\dot{L}_{\rm Edd}$ is the Eddington luminosity for $M_{\rm BH}=10^8 M_{\sun}$.}
    \label{fig:mdot}
\end{figure}

The simulations are executed in several tens of free-fall time-scale $\tau_{\rm ff,o}$ at the outer radial boundary $r_{\rm o}=1500\rs$. The mass loss rate defined as
\begin{equation}
    \Dot{M}_w (t,r=r_{\rm o})=\int_0^{80^\circ} 4\pi\rho\textbf{max}(v_r,0) r_{\rm o}^2{\rm sin}\theta \intd\theta
\end{equation}
for each model is shown in Fig.~\ref{fig:mdot}. The reason for the integral range extending to $80^\circ$ instead of $\pi/2$ is that the flow near the equatorial plane varies dramatically as turbulence. As the flow fluctuates, $80^\circ$ is a rough but tolerant approximation. It reflects more realistic outflow rate outside of the turbulent region. The mass loss rate fluctuates within one order of magnitude, which implies that the system evolves to a quasi-steady state. Thus, it is reasonable to analyse the time-averaged quantities.

\begin{figure*}
	\includegraphics[width=0.9\textwidth]{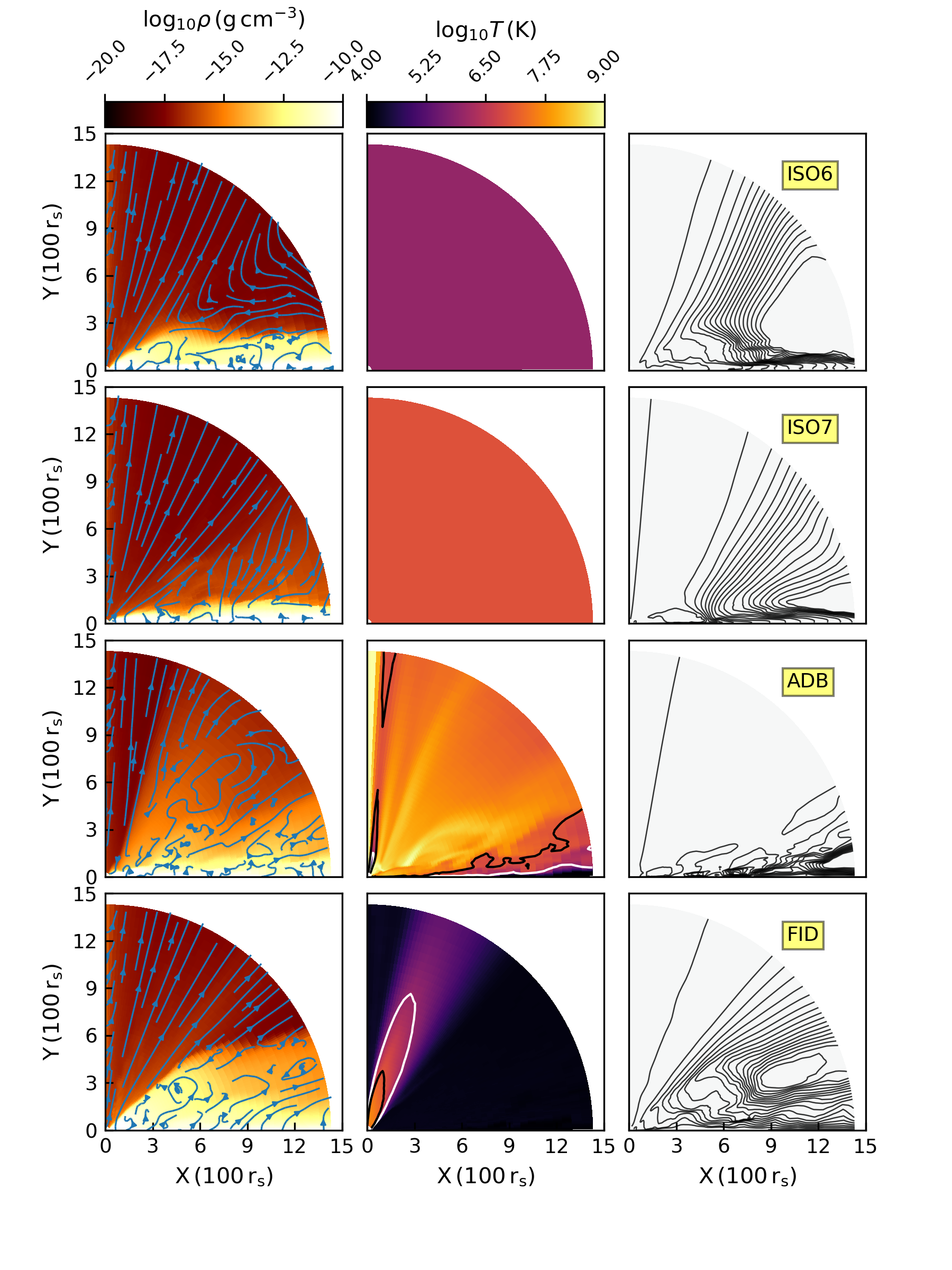}
    \caption{From left to right columns show logarithmic time-averaged density (colormap) overlaid by streamlines (blue lines with arrow), temperature (colormap) overlaid by contours with $T=10^{6,7}$K (white, black) and magnetic field lines (black contours), respectively. From top to bottom are quantities for each model labelled on the top-right corner of each panel in the last column. }
    \label{fig:timeave}
\end{figure*}

Fig.~\ref{fig:timeave} displays time-averaged quantities of each model. As the density and streamlines in Fig.~\ref{fig:timeave} show, the basic morphology in the simulated region is similar to each other in different thermodynamic models. The whole region is divided into two main parts: laminar wind close to the pole and turbulent flow close to the equator. The turbulent region is gas puffed up from the disc surface and the laminar wind is launched from the turbulent region. The fraction of each part differs in different models. 

The temperature in isothermal models (ISO6 and ISO7) and adiabatic model (ADB) is much higher than that in the fiducial model (FID) globally. The turbulent region in the former three models is more like hot corona. The compression work accounts for the high temperature of model ADB, which is a natural result of adiabatic process. However, in the fiducial model the turbulent gas resembles cold flow. It indicates that the radiative cooling is rather efficient in dense region.

\begin{figure}
	\includegraphics[width=\columnwidth]{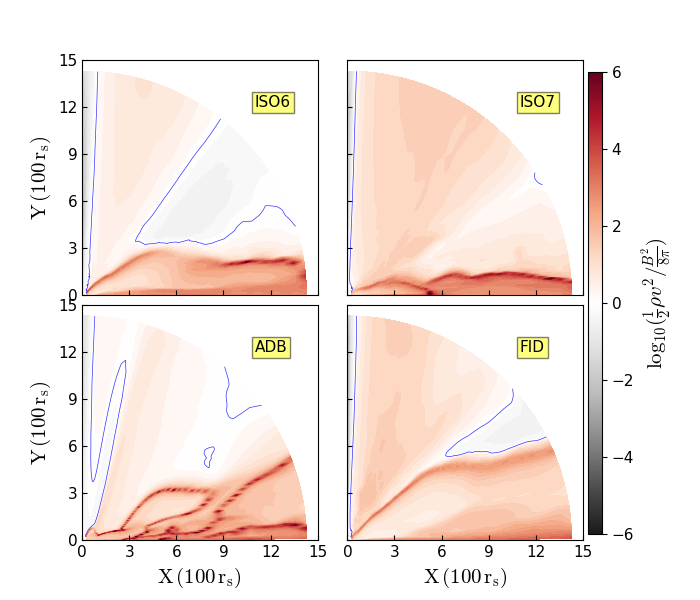}
    \caption{Colormap shows the ratio of the time-averaged kinetic energy density to the magnetic pressure for each model, blue contours denote where the ratio equals 1.}
    \label{fig:vbratio}
\end{figure}

The magnetic fields are ordered where the gas is sparse, while they are entangled where the gas is dense. The ordered large-scale field lines corresponding to outflow streamlines reflect the ideal-MHD scheme. As no physical diffusion of magnetic field is considered in the present paper, the entangled field might be adjusted by the numerical diffusion. The gas internal energy is much smaller than magnetic energy. At first glance, one would expect that the magnetic field can not be deformed. However, for a thin disc, the gas temperature is much smaller than the Virial temperature. The sound speed is much smaller than the Keplerian velocity. Even though the gas internal energy is much low, the kinetic energy of gas is much high. We compare the gas kinetic energy to the magnetic energy in Fig.~\ref{fig:vbratio}. It can been seen in our models, in most region of the computational domain, the kinetic energy is much higher than the magnetic energy. Therefore, the magnetic field can be deformed by the motions of the gas. Generally, the essential morphologies of the magnetic fields as well as the fluids are similar in models ISO6, ISO7 and FID. Model ADB has hotter and more stochastic flow than others, which leads to larger spacial fraction of turbulent region. 

\begin{figure}
    \centering
    \includegraphics[width=\columnwidth]{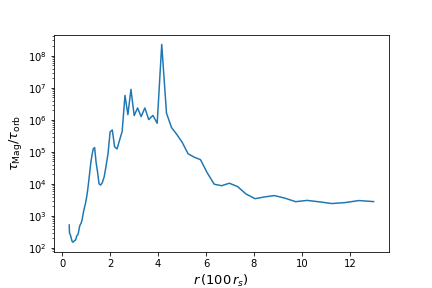}
    \caption{Timescale of angular momentum transport due to Maxwell stress on the disc in model FID, normalized by the local orbital time.}
    \label{fig:AMtrnsprt}
\end{figure}

\begin{figure}
	\includegraphics[width=\columnwidth]{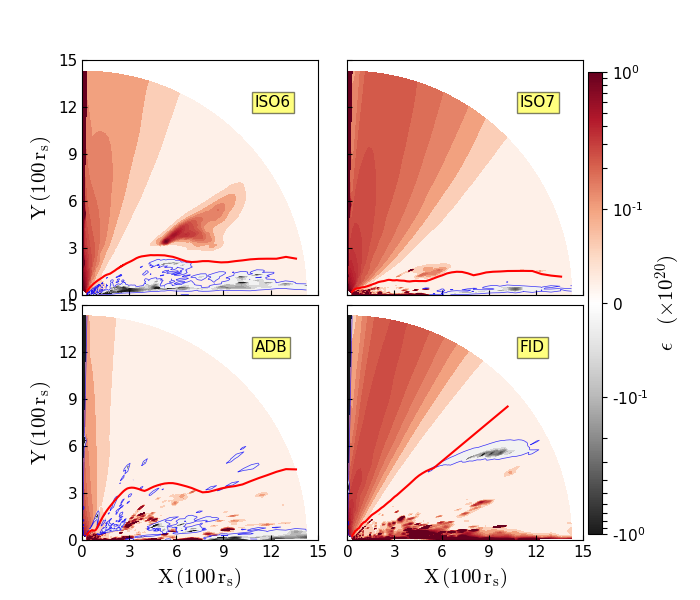}
    \caption{Colormap of the time-averaged Bernoulli parameter for each model, blue contours denote where $\epsilon=0$ and red curves denote where $\rho=10^{-15}\,\gram\cm^3$, except that the red curve for model FID is adjusted (see context in Section \ref{subsec:realwind}).}
    \label{fig:Be}
\end{figure}

\begin{figure}
	\includegraphics[width=0.8\columnwidth]{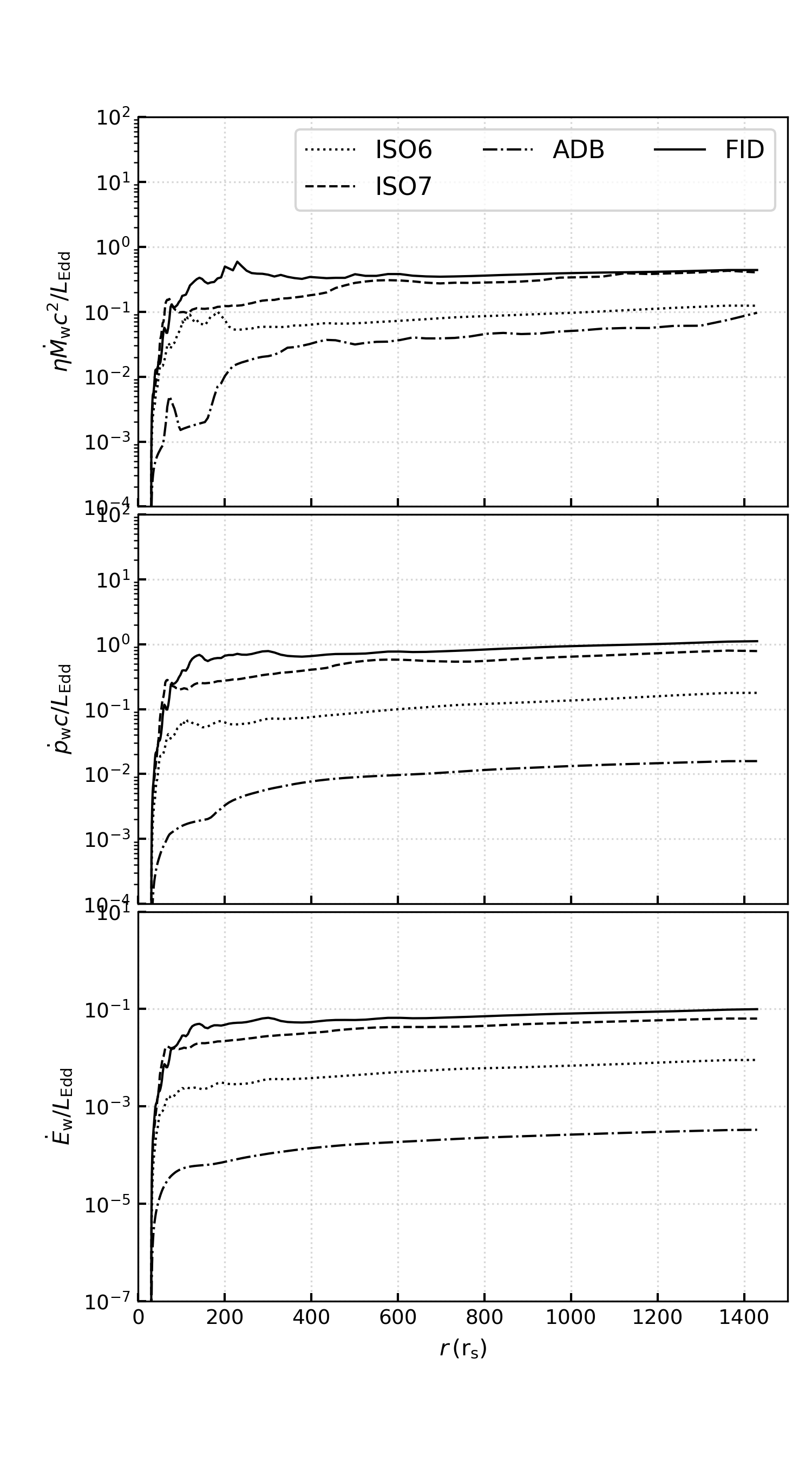}
    \caption{Radial profile of time-averaged mass flux (top panel), momentum flux (middle panel) and kinetic energy flux (bottom panel) of real winds}
    \label{fig:flux}
\end{figure}

\begin{figure}
	\includegraphics[width=0.8\columnwidth]{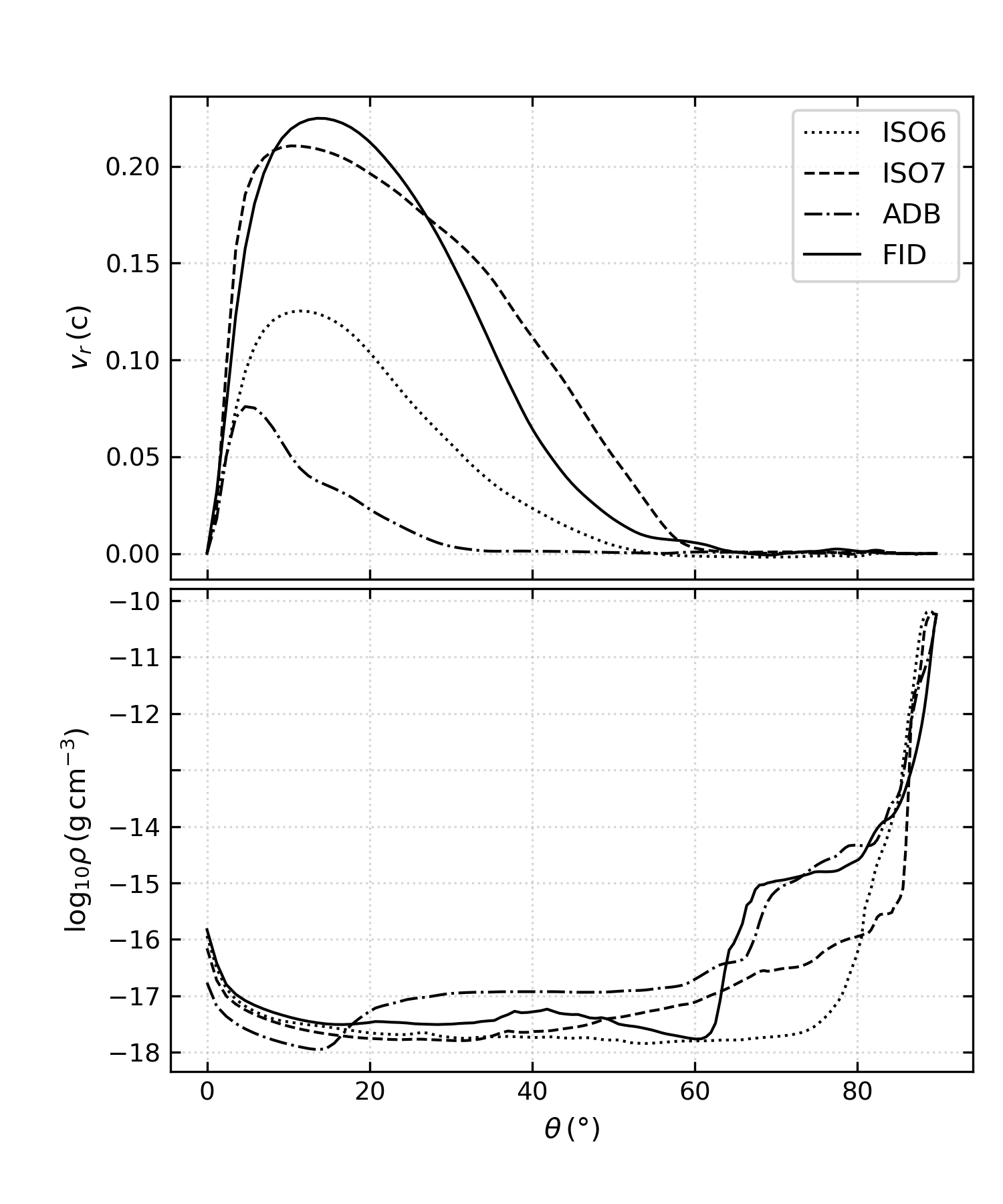}
    \caption{Angular profile of time-averaged radial velocity (upper panel) and logarithmic density (lower panel) at the outer radial boundary}
    \label{fig:angprof}
\end{figure}

The vertical magnetic field can subtract angular momentum from the disc. The timescale for the angular momentum subtraction is,  
\begin{equation}
    \tau_{\rm mag}=\frac{2\int_{0}^{H} \rho v_\phi \intd z}{2RT_{z\phi}^{\rm Max}|_{0}^{H}},
\end{equation}
where $H$ denotes the disc scale height. In Fig.~\ref{fig:AMtrnsprt}, we plot the ratio of the angular momentum subtraction timescale to the disc orbital timescale. The result shows that the angular momentum subtraction timescale is at least more than 2 orders of magnitude times larger than the orbital timescale. For most of our computational domain, the simulation time is much shorter than the angular momentum subtraction timescale, which indicates that the specific angular momentum of disc is not subtracted too much within the evolution time.

\subsection{Real wind identification}\label{subsec:realwind}
The streamlines overlaying on density in Fig.~\ref{fig:timeave} suggest that not all materials from the disc surface would continuously flow outwards. Only those materials that with positive Bernoulli constant will flow out to infinity. In this sense, to identify the real winds, we plot Bernoulli parameter calculated by
\begin{equation}\label{eqn:be_prim}
    \epsilon=\frac{1}{2}v^2+\frac{GM}{r}+h+\frac{B^2_\phi}{\rho}-\frac{v_\phi B_\phi B_p}{\rho v_p},
\end{equation}
where $h\equiv e+\frac{P}{\rho}$ is the specific enthalpy. For isothermal case, the specific enthalpy is determined by the sound speed solely $h=\frac{\gamma}{\gamma-1}c^2_{\rm s}$.

As Fig.~\ref{fig:Be} shows, the Bernoulli parameter close to the rotational axis is nearly constant with radius. It coincides with the radially laminar streamlines of winds shown in Fig.~\ref{fig:timeave}, indicating the winds are in quasi-steady state. Positive values (reddish color) represents potential outflows that have sufficient energy to surpass gravitational potential energy. The combination of the positive $\epsilon$ and laminar streamlines distinguishes the real wind from the turbulent gas which might fall back rather than flow out to the infinity. The density contour of $\rho=10^{-15}\,\gram\cm^3$ in Fig.~\ref{fig:Be} roughly separates the whole region into the turbulent flow and the laminar outflow. Thus, the position of this density contour is regarded as the wind base. For the model FID, to avoid the bulk of flow with negative $\epsilon$ at around ($900\rs,600\rs$), we constrain the density contour by substituting the polar angle outside $600\rs$ with the angle at $600\rs$ (see the right-bottom panel in Fig.~\ref{fig:Be}).

\subsection{Wind properties dependence on the thermodynamics}\label{subsec:windprop}
Fig.~\ref{fig:flux} shows radial profiles of outflow mass flux (mass loss rate), momentum flux and energy flux of real winds. They are calculated as follows,
\begin{eqnarray}
\dot{M}_w (r) = 4\pi r^2\int_0^{\theta_{\rm max}(r)}\langle\rho\rangle \textbf{max}(\langle v_r\rangle,0)  \textrm{sin}\theta \textrm{d}\theta, \label{eqn:flux1}\\
\dot{p}_w (r) = 4\pi r^2\int_0^{\theta_{\rm max}(r)}\langle\rho\rangle \textbf{max}(\langle v_r\rangle,0)^2 \textrm{sin}\theta \textrm{d}\theta, \label{eqn:flux2}\\
\dot{E}_w (r) = 4\pi r^2\int_0^{\theta_{\rm max}(r)}\frac{1}{2}\langle \rho\rangle \textbf{max}(\langle v_r\rangle,0)^3 \textrm{sin}\theta \textrm{d}\theta \label{eqn:flux3},
\end{eqnarray}
where $\langle\rho\rangle$ and $\langle v_r\rangle$ are time-averaged density and radial velocity, and $\theta_{\rm max}(r)$ is the angle of the interface between the laminar flow and the turbulent region at each radius. Quantities at each radius are integrated from the polar axis ($\theta=0$) to the wind base ($\theta_{\rm max}(r)$) which is roughly identified by the density contours demonstrated in Fig.~\ref{fig:Be} in Section \ref{subsec:realwind}.

The outflow mass flux rapidly increases within about $200\rs$ and remains nearly constant at outer radii. The model FID with cooling function gives the largest value of each flux than other models, which suggests more efficient wind production in model FID. The fluxes in model ISO6 and ISO7 are several times lower than those in model FID. The deviation of model ADB can go up to several orders of magnitude. The large deviation is consistent with the morphologies of magnetic fields and flows that implied in Fig.~\ref{fig:timeave}. Here note that the ratio of outflow rate and accretion rate exceeds unity in models ISO6, ISO7, and FID. It is not contradictory with mass conservation. In the simulation, the accretion rate is not self-consistently calculated from an evolving disc but fixed to a constant value corresponding to the assumption of a steady disc. The assumed constant density profile of the disc surface only works as gas supply. The fixed density profile can produce a corresponding accretion rate much larger than $0.1\dot{M}_{\rm Edd}$, which leads to a mass loss rate greater than $0.1\dot{M}_{\rm Edd}$.

Initially, our magnetic field has vertical component. Such a configuration is expected when the strong disc wind is emanated. The upper panel in Fig.~\ref{fig:flux} shows the radial dependence of the wind mass flux. It is shown that the outflow rate keeps increasing outside 100rs. In the fiducial model for instance, the mass flux from $100\rs$ to $200\rs$ increased from around 0.1 to 0.5 Eddington rate. Inside $100\rs$, the contribution of wind mass flux is smaller than 0.1 Eddington rate. In other words, most of the wind mass flux comes from the region outside $100\rs$. For the adiabatic model, the mass flux of wind keeps increasing from the inner boundary to the outer boundary. In this model, winds can be generated at every radii. Therefore, the assumption that the initial magnetic field has a vertical component is reasonable.

Fig.~\ref{fig:angprof} demonstrates the angular distribution of winds at the outer radial boundary. The winds in model FID can be accelerated up to $0.3 c$; the maximum velocity of winds in other models are also in the order of $0.3 c$. The winds with sufficiently large velocity have low density. They are close to the polar axis rather than the disc plane, which is a distinguished feature compared to the line-driven winds \citet{Nomura2016,NomuraOhsuga2017}. 

The adiabatic model gives the lowest velocity, mass flux and spatial fraction of winds and overestimates the temperature. It deviates the most from the fiducial model. The essential reason would be that the overheated flow suppresses the effects of magnetic fields, leading to a more stochastic flow and more tangled field lines (see the third column in Fig.~\ref{fig:timeave}). However, large-scale ordered field lines are substantial to the acceleration of winds via the Blandford \& Payne mechanism. Less powerful wind is attributed to the lack of large-scale ordered field lines.

\begin{figure}
	\includegraphics[width=\columnwidth]{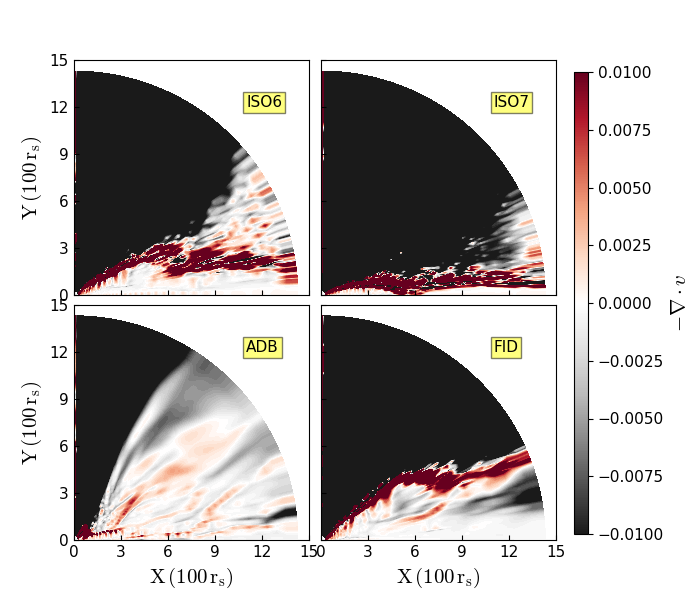}
    \caption{Negative divergence of time-averaged velocity for each model}
    \label{fig:divv}
\end{figure}

According to the first law of thermodynamics, it is apparent that the increase in the internal energy is attributed to the compression work for an adiabatic model; for isothermal models, constant internal energy constrains an artificial cooling/heating to exactly balance the compression/expansion work. To further examine the thermodynamics, we calculate the divergence of the fluid velocity to trace the compression and expansion. The results are shown in Fig.~\ref{fig:divv}. 

Fig.~\ref{fig:divv} illustrates that most compression occur in the turbulent region, especially at the wind base. For model ADB, the compression operates in almost the whole region. It confirms that compression plays a dominant role during the evolution in model ADB leading to a high temperature shown in Fig.~\ref{fig:timeave}. Compared to that in models ADB, ISO6 and ISO7, the gas within $\theta>\sim60^\circ$ in model FID is much cooler even there is compression work. It indicates the radiative cooling in the turbulent region is dominant in model FID compared to the compression work and radiative heating. 

In the wind region ($\theta<\sim60^\circ$) in all models, fluids are expanding. The expansion does negative work and cools the flow. In the model FID, compared to the turbulent region, the relatively high temperature of winds indicates that radiative heating is operative in the wind region. It also coincides with the region of high ionization parameter (see the colormap in Fig.~\ref{fig:fid_strm}). This is a natural consequence of X-ray heating. Another eminent feature in the model FID is that the temperature of winds is radially stratified, which cannot be described by the adiabatic and isothermal models.

To summarise, cooling function plays an essential role in determining the temperature of disc winds. Adiabatic assumption incorrectly overestimates the temperature. Isothermal assumption erases the local structure and underestimate the power of winds. Although the isothermal model with an appropriate temperature ($10^7$K in this work) produces similar wind dynamics to the fiducial model, this unique value of temperature cannot be provided as a priori parameter. In the following sections, we represent further detailed results of disc winds for model FID in which a cooling function is employed to mimic a more realistic case.

\subsection{Mechanisms of driving disc winds in model FID}\label{subsec:fid_mech}
\begin{figure}
        \centering
            \subcaptionbox{Time-averaged logarithmic ionization parameter $\xi$ (colormap) overlaid by selected streamlines (black curves) labelled as A and B (thick black curves) for model FID.  \label{fig:fid_strm}}{\includegraphics[width=\linewidth]{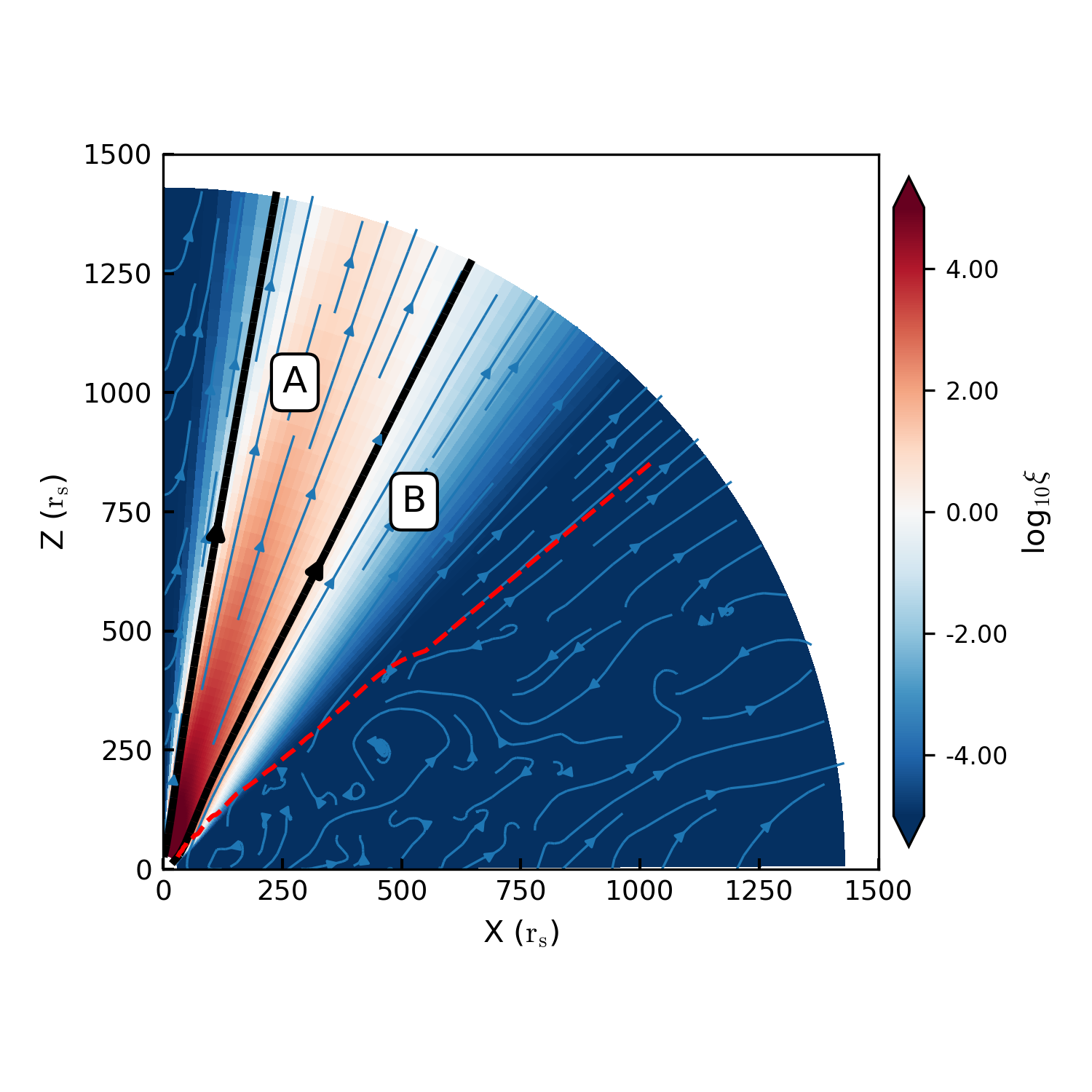}}\par 
            \subcaptionbox{The radial time-averaged acceleration along streamline A (top panel) and B (bottom panel) for model FID. The radial accelerations are centrifugal force (blue dashed line), gas pressure gradient force (green dotted line), gravity (orange solid line), magnetic pressure gradient force (yellow dash-dotted line), Lorentz force (red dash-dotted line) and total force (black solid line), respectively. All the forces are in code unit. \label{fig:fid_force}}{\includegraphics[width=\linewidth]{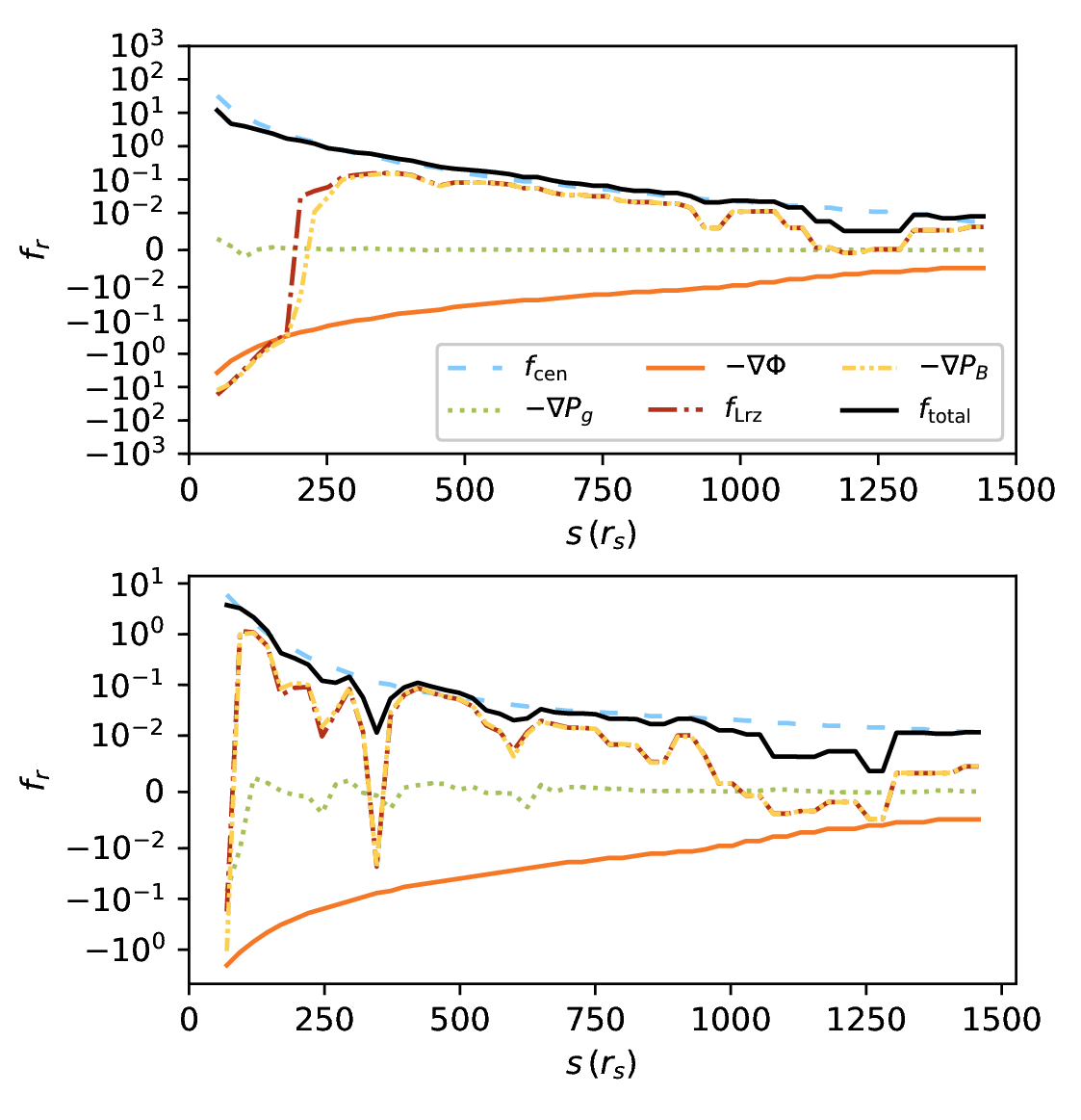}}\par 
        \label{fig:fid_dyn}
        \caption{Accelerations along streamlines for model FID}
\end{figure}

To quantitatively study the winds accelerations in model FID, we have calculated various forces per unit mass exerted along the two streamlines ``A'' and ``B'' (refer to  Fig.~\ref{fig:fid_strm}; the latter is closer to the surface of turbulent region).  The results are shown in Fig.~\ref{fig:fid_force}. The Lorentz force is
\begin{equation}(\nabla\times B) \times B = (B \cdot \nabla B)-\left(\nabla\frac{B^2}{2}\right).
\end{equation} 
The first term on the right hand side of this equation is the magnetic tension force and the second term is the magnetic pressure gradient force. The magnetic tension force is negligibly small, so only the Lorentz force and the magnetic pressure gradient force are presented.

Along both streamlines, the winds are accelerated by both the centrifugal and magnetic pressure gradient forces. The centrifugal force is slightly stronger than the  magnetic pressure gradient force. The gas pressure gradient force is negligible as expected. It is a natural result for the cold accretion disc, where gas temperature is significantly smaller than the Virial temperature. The acceleration along the streamline B reminds us the wind acceleration model in \citet{BlandfordPayne1982}.  We find that the magnetic field lines do have an appropriate inclination angle with the rotational axis (recall the magnetic field lines in Fig.~\ref{fig:timeave}), as required by the Blandford \& Payne mechanism to work.

\subsection{Comparison to observations: UFOs}\label{subsec:fid_obs}
\begin{figure}
    \centering
    \includegraphics[width=\columnwidth]{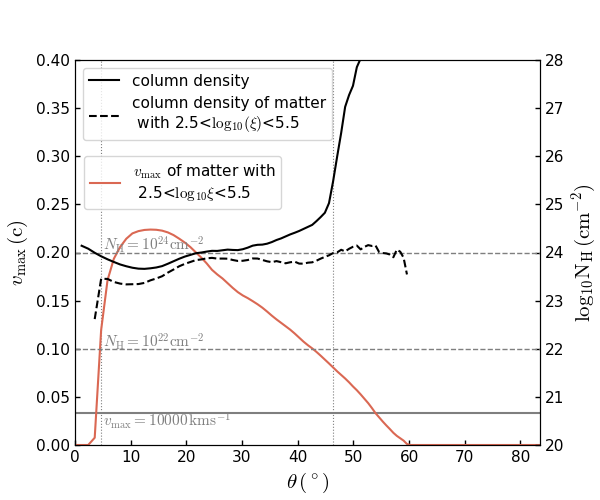}
    \caption{Column density (black solid line), column density of gas with $2.5 < {\rm log \xi} < 5.5$ (black dashed line), and maximum velocity of the gas with $2.5 < {\rm log \xi} < 5.5$ (red line). The horizontal axis is viewing angle (or $\theta$ angle). The vertical axis on the left-hand side shows the value of velocity (red line). The vertical axis on the right-hand side shows the value of column density (black solid and black dashed lines). The horizontal solid line marks the velocity of $10000{\rm km/s}$. The horizontal dashed lines mark column density of $10^{22} {\rm cm^{-2}}$ and $10^{24} {\rm cm^{-2}}$. The vertical dotted lines mark $\theta=4.7^{\circ}$ and $45.6^{\circ}$ where the disc winds satisfy the UFO conditions in between.}
    \label{fig:columndvmax_ufo}
\end{figure}

The disc winds produced in our simulations have inhomogeneous properties and extensive range of the ionization parameter. Not all winds meet the conditions of the observed UFOs. To compare the observable variables, we exploit the constraints of the observed features from \citet[]{Gofford2015} on the disc winds. Generally, UFOs have velocity higher than $10^4~{\rm km~s^{-1}}$, ionization parameter in the range $2.5 < {\rm log \xi} < 5.5$ and column density in the range $22 < {\rm log N_H} < 24$. The angular distribution of these UFOs is shown in Fig.~\ref{fig:columndvmax_ufo}. Similarly, the UFOs locate close to the pole rather than the equator. They cover the region between $\theta=4.7^{\circ}$ and $45.6^{\circ}$. The corresponding solid angle $\Omega\approx 3.85$; the fraction of $\Omega$ to a whole sphere, $\Omega/4\pi\approx 30.6$ per cent. It is close to the lower limit of the fraction of UFO in AGNs \citep[][]{Gofford2015}.

\begin{figure}
    \centering
    \includegraphics[width=\columnwidth]{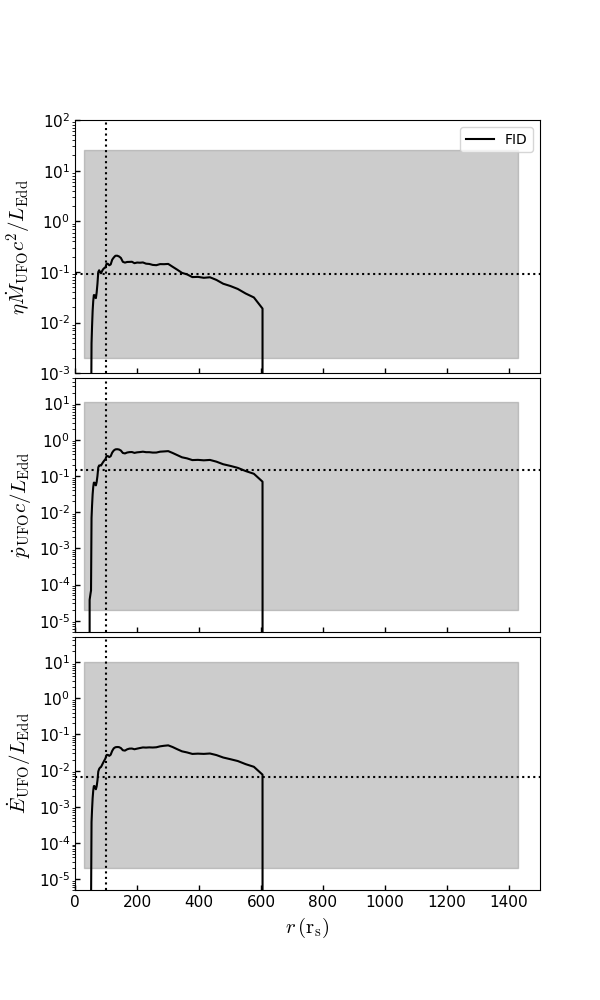}
    \caption{Mass flux (top), momentum flux (middle) and kinetic power (bottom) of UFOs in model FID (black curve). All of the properties of UFOs in the sample of \citet{Gofford2015} are located inside the shade region. In other words, the upper and lower boundaries of the shade region correspond to the maximum and minimum values of properties of the observed UFOs. The horizontal dotted line is the averaged values from observations \citep{Gofford2015}.  The vertical dotted line denotes the innermost radii of oberved UFOs \citep{Gofford2015}, i.e. about $100\rs$. }
    \label{fig:flux_ufo}
\end{figure}

To obtain the mass flux, momentum flux and kinetic power of UFOs, we integrate the Equation (\ref{eqn:flux1})-(\ref{eqn:flux3}) over the flows which meet the above conditions. The comparison of the UFOs found in model FID to observations is shown in Fig.~\ref{fig:flux_ufo}. The cut-off at $600\rs$ implies that in model FID no wind meets the condition of UFOs outside $600\rs$. The properties of UFOs within that radius are well in the range given by observations and close to the averaged values.

Combining the ionization parameter (colormap in Fig.~\ref{fig:fid_strm}) and the temperature (second column in Fig.~\ref{fig:timeave}) of model FID, the reason of the missing UFOs at large radii is the low ionization state. However, it has been reported that the UFOs can exist within $10^2-10^4\rs$ \citep{Gofford2015}. Thus, some physics might be missing in MHD disc winds models. A plausible conjecture for the missing physics would be line force. As the winds are sparse and cool, line force might operate on the winds to change their properties.

\section{Parameter dependence of UFOs}

\subsection{Dependence on midplane gas density and X-ray flux}
\begin{figure}
    \centering
    \includegraphics[width=\columnwidth]{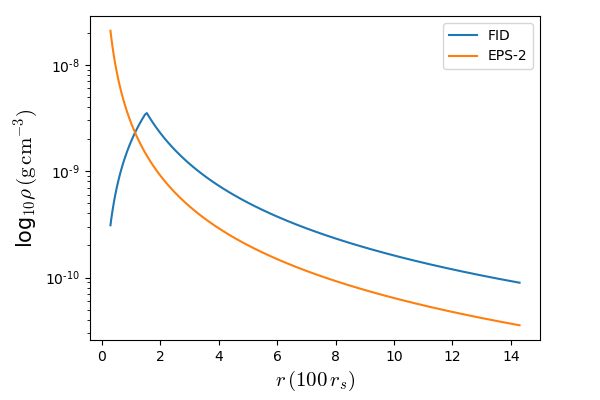}
    \caption{Density profile on midplane (disc surface) derived from Equation (11) with $\varepsilon=0.1$ (blue) and $\varepsilon=0.01$ (yellow).}
    \label{fig:discdensity}
\end{figure}

In our fiducial model, the gas density at the midplane is set corresponding to that of a thin disc which has Eddington ratio of 0.1. We do a test simulation EPS-2. In this simulation, the midplane density is also obtained by Equation (11) by setting $\varepsilon$ to 0.01. Also, in the test simulation, the X-ray luminosity is 10 times lower than that in the fiducial model. In Fig.~\ref{fig:discdensity}, we plot the midplane density in our fiducial model (blue line) and test simulation EPS-2 (yellow line). It can be seen that inside 200$r_s$, where the UFOs are launched, the midplane density in model EPS-2 is much higher than that in our fiducial model. Outside 200$r_s$, the density in model EPS-2 is much lower.

Fig.~\ref{fig:parflux_ufo}  (black lines) shows the UFOs in model EPS-2. The UFOs are present only inside 100$r_s$. The reason of the absence of UFOs outside 100$r_s$ is as follows. We find that the wind density in the region $\theta < 45^\circ$ in model EPS-2 is comparable to that in the fiducial model. However, the X-ray flux in model EPS-2 is 10 times lower. Thus, the winds at large radii in model EPS-2 have too low ionization parameter to satisfy the conditions of the observed UFOs. Both the mass flux and kinetic power of UFOs in model EPS-2 are several times lower than those in the fiducial model. The reason is the same that in model EPS-2, the ionization parameter is relatively low. Therefore, the winds which satisfy the conditions of UFOs are reduced.

Observations show that the UFOs are located at the region of 100$r_s$ -- 10000$r_s$ \citep{Gofford2015}. In model EPS-2, the UFOs are located inside 100$r_s$. Therefore, it is not consistent with observations.  We also find that the maximum velocity of UFOs in model EPS-2 is quite similar as that in the fiducial model. This is not consistent with observation which show that the UFOs velocity is faster for higher luminosity.

\subsection{Dependence of UFOs on magnetic field strength}
\begin{figure}
    \centering
    \includegraphics[width=\columnwidth]{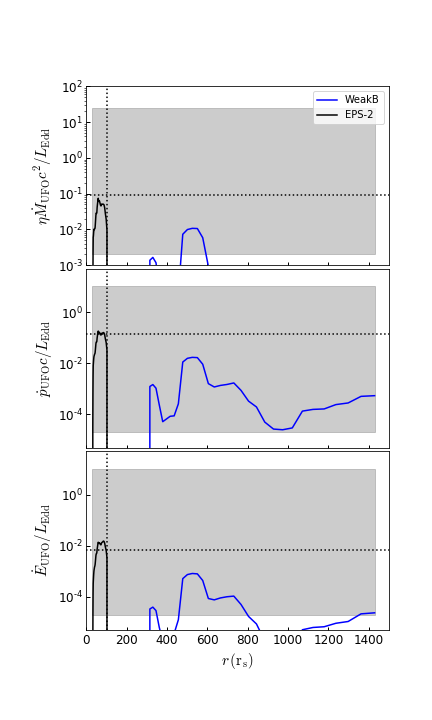}
    \caption{Same as Fig.~\ref{fig:flux_ufo} but for model EPS-2 with $\varepsilon=0.01$ (black) and model WeakB with a weak B-field case (blue)}
    \label{fig:parflux_ufo}
\end{figure}

In our fiducial model, when the quasi-steady state is achieved, at the midplane, the plasma beta $\beta \sim 10^{-4}$. We perform a test simulation to study the dependence of UFOs on the magnetic field strength. In our test simulation, when the quasi-steady state is achieved, the midplane plasma beta $\beta \sim 10^{-2}$. The magnetic field strength in the test simulation is much lower. The mass flux, momentum flux and kinetic power of UFOs are plotted in Fig.~\ref{fig:parflux_ufo} (blue curve). Comparing figures \ref{fig:flux_ufo} with \ref{fig:parflux_ufo}, we can see that, in the test simulations, the mass flux of UFOs is one order of magnitude smaller than that in the fiducial model. The kinetic power of UFOs in the test simulation is more than 1 order of magnitude smaller than that in the fiducial model. In future, further complete parameter survey is necessary to draw a quantitatively solid conclusion of the dependence of UFOs on magnetic field strength.

%%%%%%%%%%%%%%%%%%%%% CONCLUSIONS %%%%%%%%%%%%%%%%
\section{Summary}\label{sec:sum}
We have studied the difference of wind properties by considering different thermodynamics: isothermal, adiabatic and radiative. The wind properties strongly depend on the treatment of thermodynamics. The isothermal and adiabatic models underestimates the wind power and overestimates the wind temperature. Additionally, the isothermal and adiabatic assumptions erase the local structure of disc winds. We suggest future work consider energy equation with radiative cooling and heating to obtain accurate wind properties. In this way the wind model can be more appropriate to directly compared to the observations.

The fiducial model with radiative cooling and heating can produce powerful radially stratified winds. The winds inside $600\rs$ are consistent with the observations of UFOs, while no UFOs are found outside $600\rs$. The lack of UFOs with appropriate ionization parameter, column density and velocity at large radii implies that extra physics need to be considered. In future, seeking these extra physics would be meaningful to the interpretation of the observed winds from a thin accretion disc.

\section*{Acknowledgements}
This work has made use of the High Performance Computing Resource in the Core Facility for Advanced Research Computing at Shanghai Astronomical Observatory. This work is supported in part by the Natural Science Foundation of China (grants 12173065, 12133008, 12192220, and 12192223) and the science research grants from the China Manned Space Project (No. CMS-CSST-2021-B02). W.W. thanks Jiawen Li and Can Cui for the helpful discussions. The authors also would like to thank the anonymous referee for the constructive and insightful comments on the manuscript. This research has made use of NASA's Astrophysics Data System Bibliographic Services.

\section*{DATA AVAILABILITY}
The data underlying this article will be shared on reasonable request to the corresponding author.


\bibliographystyle{mnras}
\bibliography{ms} % if your bibtex file is called example.bib

% Don't change these lines
\bsp	% typesetting comment
\label{lastpage}
\end{document}